\documentclass[%
 reprint,
amsmath, amssymb,
aps,
]{revtex4-1}
\usepackage{graphicx}
\usepackage{makecell}
\usepackage[normalem]{ulem} 
\usepackage{dcolumn}
\usepackage{bm}

\usepackage{xcolor}
\usepackage{graphicx}
\usepackage{epsfig}
\usepackage{slashed}
\usepackage{tabu,multirow}
\usepackage[unicode=true,bookmarks=false,breaklinks=false,pdfborder={0 0 1},colorlinks=true]
 {hyperref}
\hypersetup{
 citecolor=blue,linkcolor=blue,urlcolor=blue}

\begin{document}


\title{A holographic bottom-up approach to $\Sigma$ baryons}

\author{Xi Guo }
\email{Gxiguoxi@163.com}
\affiliation{School of Nuclear Science  and Technology\\
 University of South China\\
 Hengyang, China\\
 No 28, West Changsheng Road, Hengyang City, Hunan Province, China.}

 \author{Miguel Angel Martin Contreras}
\email{miguelangel.martin@usc.edu.cn}
\affiliation{School of Nuclear Science  and Technology\\
 University of South China\\
 Hengyang, China\\
 No 28, West Changsheng Road, Hengyang City, Hunan Province, China.}

\author{Xun Chen}
\email{chenxunhep@qq.com}
\affiliation{School of Nuclear Science  and Technology\\
 University of South China\\
 Hengyang, China\\
 No 28, West Changsheng Road, Hengyang City, Hunan Province, China.}

\author{Dong Xiang}
\email{xiangdong@usc.edu.cn}
\affiliation{School of Nuclear Science  and Technology\\
 University of South China\\
 Hengyang, China\\
 No 28, West Changsheng Road, Hengyang City, Hunan Province, China.}

\begin{abstract}
In this work, we discuss the description of neutral $\Sigma$ baryons with $I(J^P)=1(1/2^+)$ and $I(J^P)=1(3/2^+)$ using two bottom-up approaches: the deformed background and the static dilaton. In both models, we consider a non-linear Regge trajectory extension motivated by the \emph{strange} nature that $\Sigma$ baryons have. We found that both models describe these systems with an RMS error smaller than 10 $\%$. We also perform a configurational entropy calculation in both models to discuss hadronic stability.

\end{abstract}

\maketitle

\section{Introduction}

Describing baryons in the AdS/QCD bottom approach follows the same path as meson states. We start with an action for bulk fields dual to baryons at the conformal boundary. However, confinement is performed slightly differently from the meson scenario. Confinement is understood in these bottom-up scenarios as a \emph{localization process}, where the non-normalizable bulk modes become normalizable by adding \emph{properly} an energy scale. This energy scale will later fix the Regge slope. This localization mechanism can be done by two possible alternatives: cutting off or deforming the AdS geometry, i.e., breaking the conformal invariance in the bulk.

The cutoff of the space leads to the so-called hardwall \cite{Boschi-Filho:2002wdj, Erlich:2005qh,deTeramond:2005su} and softwall models \cite{Karch:2006pv}. The former is achieved by placing a D-brane into the AdS space. The locus of the brane defines the energy scale as $\Lambda_\text{QCD} = 1/z_c \propto M_0$, where $M_0$ is the lightest hadron on the Regge trajectory. The mass spectrum usually behaves as  $M_n \propto n$, which is unexpected from light-unflavored hadron phenomenology. Quantum mechanically, this model behaves qualitatively like an \emph{infinite square well}. The softwall model case includes a dilaton field that can be static or dynamically generated \cite{Batell:2008zm}, smoothly breaking conformal invariance. The net effect of this dilaton is the emergence of \emph{linear confinement} \cite{Karch:2006pv} when considering quadratic dilaton profiles, leading to Regge-like behavior $M_n^2\propto n$, with $n$ defined as the excitation number, expected for unflavored light. Quantum mechanically speaking, these holographic potentials defined with quadratic dilatons behave as harmonic oscillators in the polar plane.

The softwall model proposal has proved to be successful in describing the spectroscopy of light-unflavored hadrons \cite{BallonBayona:2007vp, Vega:2010ns, Branz:2010ub, Afonin:2010fr, Colangelo:2008us, Colangelo:2010pe, BallonBayona:2007qr, Gutsche:2011vb, Li:2012ay, Li:2013oda, Li:2016gfn, Vega:2020ctz},  form factors \cite{Abidin:2008ku, Abidin:2008hn, Abidin:2009hr} structure functions \cite{Watanabe:2013spa}, deep inelastic scattering \cite{BallonBayona:2007qr, Braga:2011wa}. It has been extended to the light-front case \cite{Brodsky:2007hb, deTeramond:2008ht, Brodsky:2008pf}, where it has opened an active research area for wave functions \cite{Brodsky:2006uqa, Vega:2009zb}, form factors \cite{Sufian:2016vso, Chakrabarti:2020kdc},  and spectroscopy \cite{Ahmady:2021yzh}.

However, despite all its success in the light sector, it could not properly describe heavy mesons since their Regge trajectories are not linear. From the Bethe-Salpeter perspective, by including the quark constituent mass linearity is lifted, i.e., $M_n^2\propto n^{\nu}$, where $\nu$ depends on the constituent mass \cite{Chen:2018bbr, Chen:2018hnx, Chen:2018nnr}. Along with this issue comes the mesonic decay constants issue. For bottom-up models, the decay constants do not match the phenomenological expected behavior; mesonic decay constants (in units of MeV) are expected to decrease with the excitation number. For hardwall, they increase, and for softwall, they are degenerate. Attempts to preserve quadratic behavior and obtain acceptable phenomenological decays have been made in the past \cite{Grigoryan:2010pj, Braga:2015jca}. However, as pointed out in Ref. \cite{MartinContreras:2019kah}, decay constants also depend on the low-$z$ behavior of the dilaton field. This situation is precisely the scenario proposed in Ref. \cite{Braga:2017bml} for heavy vector quarkonia.

Regarding the heavy-spectroscopy issue, Ref. \cite{MartinContreras:2020cyg} has extended the softwall model to include non-linear Regge trajectories by promoting the quadratic dilaton $\kappa^2\,z^3$ to be \emph{deformed} $(\kappa\,z)^{2-\alpha}$ by a parameter that accounts constituent mass effects. This proposal was also successful in describing non-$q\bar{q}$ states such as hadroquarkonium, hadronic molecules, or hybrid mesons, as in Ref. \cite{MartinContreras:2023oqs}, where configurational entropy is used as a tool to test the feasibility of the proposed holographic structures.

Ref. \cite{Andreev:2006vy} originally proposed the deformation of the AdS space in the context of gauge/string duality applied to describe the OPE expansion for the two-point function. In this work, the author proves how quadratic deformations in the AdS$_5$ sector lead to Regge-like behavior. This observation leads to Ref. \cite{Forkel:2007cm}, where the authors extended this idea to compute Regge trajectories for mesons and baryons. Later, the geometric deformation was used in Ref. \cite{FolcoCapossoli:2019imm} to describe glueballs and light baryons. These works have in common that deformations in the AdS geometry induce locality by transforming bulk modes into normalizable ones, realizing confinement. This fact is translated into the emergence of confining terms in the holographic potential. In this sense, deformations and softwall models are equivalent. However, the analytical behavior of eigenmodes is completely different. In this framework, it has been possible to describe proton structure functions \cite{FolcoCapossoli:2020pks}, electromagnetic form factors for nucleons \cite{Contreras:2021epz} and pions \cite{MartinContreras:2021yfz}.

Another tool has gained importance with the holographic approaches developed to describe hadrons, the \emph{configurational entropy}. The original proposal \cite{Gleiser:2011di, Gleiser:2012tu, Gleiser:2013mga} addresses the connection between the information and physical solutions for a given system. This idea is connected to how the energy is localized in such solutions. This localization of energy is related to the emergence of order structures. Thus, the configurational entropy (CE) can be understood as a \emph{entropic measure} of how system constituents are organized in space. Holographically, CE has been extended in works as \cite{Braga:2016wzx, Braga:2017fsb} to describe black hole stability in AdS and heavy quarkonia. In particular, the authors found the connection between decay constants and configurational entropy. When the former decreases with the excitation number, the latter increases. This observation can be considered an insight into hadronic stability \emph{via} holographic tools. In this line of research, several works have enriched the literature, as indicated by Refs. \cite{daRocha:2021ntm, Barreto:2022ohl, MartinContreras:2022lxl, Zhao:2023yry, MartinContreras:2023eft}.

In this work, we will consider the approach to heavy baryons using the deformed dilaton proposal in both softwall and deformed geometry models. We will also consider configurational entropy as a tool to test hadronic stability in these models.

This work is organized as follows. In Section \ref{sec-2}, we summarize the bottom-up description of baryons. In Section \ref{sec-3}, we use geometric deformations and static dilaton in the non-linear Regge trajectory context to describe $\Sigma$ baryons. Section \ref{sec-4} has a detailed calculation of the configurational entropy for these fermionic systems. And finally, in Section \ref{sec-5}, we present our conclusions.

\section{Holographic approach to baryons}\label{sec-2}
Let us consider the AdS$_5$ space defined by the Poincarè line element as

\begin{equation}
    dS^2=e^{2\,A(z)}\,\left[dz^2+\eta_{\mu\nu}\,dx^\mu\,dx^\nu\right],
\end{equation}

\noindent where the warp factor is defined as $A(z)=\log(R/z)+h(z)$, with $R$ the radius of the AdS curvature. The function $h(z)\in \mathcal{C}^{\infty}$ is the \emph{geometric deformation} that, for the softwall model case, will be fixed as zero. We will use Latin indices for five-dimensional bulk objects and Greek indices for four-dimensional boundary objects.

Baryons, as it is standard in AdS/QCD, are described by bulk fermionic fields. However, it is not a straightforward task compared to mesons, where the effect of a dilaton field (static or dynamically generated) enters directly into the holographic potential due to the coupling of the dilaton field with bulk fields dual to mesons. In the case of baryons, the dilaton field is factorized out from the equation of motion. Thus, different mechanisms have to be considered to model these states.   This situation is avoided when geometric deformations are considered since the confinement information is condensed in the warp factor. The main objection now arises because the background is flavor-dependent.

We will describe fermionic fields in AdS backgrounds with deformation and dilaton fields, following the prescription defined in \cite{Gutsche:2011vb, FolcoCapossoli:2019imm}.

Let us focus on baryons with dilaton fields. Refs. \cite{Vega:2008te, Gutsche:2011vb} pointed out, the dilaton field can be introduced as  an \emph{anomalous dimension}  that modifies the fermion bulk mass

\begin{equation}
\tilde{M}_5(z)=M_5+\frac{\Phi(z)}{R}.
\label{eq:m5D}
\end{equation}

This modification ensures that the bulk modes become normalizable when considering dilaton-based models.

\subsection{Spin 1/2 baryons}

For $1/2$ baryons, the bulk action is written in the standard Dirac form as follows:

\begin{equation}\label{Dirac-ac}
    I=\frac{1}{\mathcal{K}}\int{d^5x\,\sqrt{-g}\left[\frac{1}{2}\bar{\psi}\,\Gamma^r\,\overset{\leftrightarrow}{D}_r\,\psi-M_5(z)\,\bar{\psi}\,\psi\right]},
\end{equation}

\noindent with $\Gamma^r=e^r_a\,\gamma^a$ as the Dirac gamma matrices in curved space, $\mathcal{K}$ is a constant fixing the units in the action, and \emph{covariant spin-connected derivative operator} $D_m$ is defined as

\begin{equation}
D_m\,\psi = \partial_m\,\psi+\frac{1}{4}\omega^{ab}_m\,\sigma_{ab}\,\psi,
\end{equation}

\noindent where $\omega^{ab}_m$ is the spin connection, $\sigma^{ab}$ is the flat gamma matrix commutator. The equations of motion for these fields read as follows.

\begin{eqnarray}
   \left(\Gamma^m\, \overset{\rightarrow}{D}_m-M_5\right)\,\psi(z,x^\mu)&=&0,\\
  \bar{\psi}(z,x^\mu) \left(\Gamma^m\, \overset{\leftarrow}{D}_m+M_5\right)&=&0.
\end{eqnarray}

For AdS$_5$, the frame field is  $e^a_m=\delta^a_m\,e^{A(z)}$, where Latin indices $a,b,c,\ldots$ denote the flat frame indices. Thus, for the spin-affine connection, the non-zero components are

\begin{equation}
    \omega^{5\,b}_\mu=-A'\left(z\right)\,\delta^b_\mu.
\end{equation}

Therefore, the Dirac equation for the bulk spinor $\psi(z,x^\mu)$ can be written as

\begin{equation}
\left[ \gamma^{5}\,\partial_z+ \eta^{\mu \nu}\,\gamma_\mu\,\partial_\nu+2\,A'\left(z\right)\gamma_{5}-M_{5}(z)\,e^{A\left(z\right)}\right]\psi\left(z,x^\mu\right)=0.
\end{equation}

A similar expression can be found for the adjoint bulk spinor $\bar{\psi}(z,x^\mu)$. Next, we introduce the chiral components for the bulk spinors as

\begin{equation}
    \psi(z,x^\mu)=\psi_L(z,x^\mu)+\psi_R(z,x^\mu).
\end{equation}

Using this definition and after transforming to the Fourier space, the Dirac equation is written as

\begin{multline}
\left[\mp\partial_z\mp2\,A'\left(z\right)-M_{5}(z)\,e^{A(z)}\right]\psi_{L/R}\\
+m\,\psi_{R/L}=0,
\end{multline}
\begin{multline}
\left[\partial_z+2\,A'\left(z\right)\pm M_{5}(z)\,e^{A(z)}\right]\psi_{L/R}\\
\mp m\,\psi_{R/L}=0.
\end{multline}

In the last equation, we used the boundary Dirac equation in Fourier space to introduce the baryonic mass $m$.

To decouple these equations, we take the second derivative, and after some algebra, we obtain the \emph{Sturm-Louville} form of the Dirac equation:

\begin{multline}\label{sturm-Lioville-Fermion}
\psi''_\pm  +3\,A'\,\psi'_\pm+\left\{4\,A'^2+2\,A''\right.\\
\left.\pm \left[M_{5}(z)\,A'+M_{5}'(z)\right]\,\,e^A-M_{5}^2(z)\,e^{2\,A}\,\right\}\psi_\pm\\
+m^2\,\psi_\pm=0.
\end{multline}

At this step, confinement emerges from the holographic potential $V(z)$. This potential is defined using the \emph{Boguliobov transformation}

\begin{equation}
\psi_{L/R}\left(z\right)=e^{-2\,A\left(z\right)}\,\phi_{L/R}\left(z\right),
\end{equation}

\noindent we obtain the \emph{Schrodinger-like form} of the bulk equations of motion:

\begin{equation}\label{Fermion-Schr}
-\phi_{L/R}''+V(z)\,\phi_{L/R}=m_n^2\,\phi_{L/R},
\end{equation}

\noindent with $p^2=-m_n^2$ and for the potential has the following structure

\begin{equation}\label{fermion-pot-sch}
 V(z)=M_{5}^2(z)\,e^{2\,A(z)}\mp \left[M_{5}(z)\,A'(z)+M_{5}'(z)\right]\,e^{A(z)}.
\end{equation}

The eigenvalues of this potential will correspond to the spin $1/2$ baryon masses at the boundary. A similar behavior is found for the adjoint spinor solutions. Both left and right solutions have the same eigenvalue mass $m_n^2$. Thus, following \cite{FolcoCapossoli:2019imm}, we choose left movers to be dual to baryons at the boundary. The last ingredient we must fix is the bulk mass to define the baryonic identity. We will discuss this topic in the next section.

In the next sections, we will discuss applying the non-quadratic dilaton and the deformed geometry in this formalism.

\subsection{Spin 3/2 baryons}

Let us consider spin 3/2 baryons defined using a Rarita-Schwinger bulk field. The bulk action in this case is given by \cite{Gutsche:2011vb}

\begin{multline}\label{rarita}
I=-\frac{1}{2\,\mathcal{K}}\int{d^5x\,\sqrt{-g}\,g^{mn}}\times\\
\left[\bar{\psi}_{m}\,\Gamma^r\,\overset{\leftrightarrow}{\nabla}_r\,\psi_{n}-M_5(z)\,\bar{\psi}_{m}\,\psi_{n}\right],
\end{multline}

\noindent where $\psi_m(z,x^\mu)$ is a bulk vector spinor, and the covariant derivative is defined as

\begin{equation}
\nabla_m\,\psi_{n}=D_m\,\psi_{n}-\Gamma^r_{m n}\,\psi_{r},
\end{equation}

\noindent with $\Gamma^r_{mn}$ is the \emph{Levi-Civita affine connection} in AdS$_5$, that has the non-zero components given by

\begin{eqnarray}
\Gamma^z_{zz}&=&A'\left(z\right),\\
\Gamma^z_{\mu\nu}&=&-A'\left(z\right)\,\eta_{\mu\nu},\\
\Gamma^\mu_{z\nu}&=&A'\left(z\right)\,\delta^\mu_\nu.
\end{eqnarray}

From the action principle \eqref{rarita}, we obtain the following bulk equations of motion

\begin{equation}
\left[g^{nm}\,\Gamma_m\,\nabla_n-M_{5}(z)\right]\,\psi_{m}=0.
\end{equation}

Let us consider the gauge fixing. Since no holographic information should be explicitly written at the boundary, i.e., no dependence on the holographic coordinate $z$ is expected, we impose

\begin{equation}
 \psi_{z}(z,x^\mu)=0.
\end{equation}

As a consequence, we will have the following set of \emph{transverse conditions}:

\begin{eqnarray}\label{fermion-gauge-2}
\Gamma^m\,\psi_{m}(z\,x^\mu)&=&0,\\   \label{fermion-gauge-3}
g^{mn}\,\nabla_m\,\psi_{n}(z,x^\mu)&=&0.
\end{eqnarray}

The condition \eqref{fermion-gauge-3} follows considering the product of the antisymmetric products of the Dirac matrices $\Gamma^{mnr}$with the symmetric spinor tensor field, i.e.

\begin{eqnarray}
\Gamma^{mnr}\,\nabla_m\,\psi_{r}+M_{5}(z)\,\Gamma^{mn}\,\psi_{n}=0,
\end{eqnarray}

\noindent which is equivalent to the Dirac equation. Using the first transverse condition will lead to the second one \eqref{fermion-gauge-3}.

After the gauge fixing process, and using the expressions for the covariant derivative in AdS$_5$, we write the equations of motion for the bulk vector spinor as

\begin{equation}\label{High-SF}
 \left[\gamma^{5}\,\partial_z+\gamma^\mu\,\partial_\mu+2\,A'\left(z\right)\,\gamma_{5}-M_{5}(z)\,e^{A\left(z\right)}\right]  \psi_{m}=0,
\end{equation}

\noindent which is the equation for spin 1/2 bulk fermions. The main difference is the bulk mass $M_5(z)$ since the operators that define spin 3/2 baryons have a different scaling dimension than the spin 1/2 ones.

Therefore, following the same procedure exposed for spin 1/2 bulk spinor, we obtain the \emph{Schrodinger-like equation of motion} as

\begin{equation}
-\phi_{L/R}''+U\left(z\right)\,\phi_{L/R}=m_n^2\,\phi_{L/R},
 \end{equation}

\noindent with the holographic potential $U(z)$ defined as

\begin{equation}
 U(z)=M_{5}^2(z)\,e^{2\,A(z)}\mp \left[M_{5}(z)\,A'(z)+M_{5}'(z)\right]\,e^{A(z)}.
\end{equation}

As in the spin 1/2 case, we consider baryons defined by left vector spinors. In the next sections, we will solve the eigenvalue problem for the non-quadratic and deformed geometry models.

\section{A$\rm \bf d$S/QCD applied to $\Sigma$ baryons}\label{sec-3}
This section will apply the above-mentioned machinery to describe the $\Sigma$ baryons spectrum.

These $\Sigma$ baryons were initially observed in cosmic ray experiments during the 1950s and have since been extensively studied in particle accelerators. These states are compounded by a pair of light quarks with an $s$ quark. They play a crucial role in understanding the strong force that binds quarks and the composition of protons and neutrons within atomic nuclei.

In particular we will focus on the neutral $\Sigma$ baryons, with $I(J^P)=1(1/2^+)$ and $I(J^P)=1(3/2^+)$. We summarize the experimental masses in Tables \ref{table:table1} and \ref{table:table2}. Neutral $\Sigma$ baryons, aside from the $\Sigma^0(1/2)$, decay into neutral kaons mostly. For the $\Sigma^0$ baryon, the dominant decay is $\Lambda\,e^+\,e^-$.

In previous studies (see \cite{Chen:2023web} and references therein), it has been found that the Regge trajectory depends on the constituent mass of quarks. Thus, if a hadron contains an $s$ or a heavy quark, the linearity of the trajectory ceases. Motivated by this observation,  we employ the non-quadratic softwall model \cite{MartinContreras:2020cyg} to address the neutral $\Sigma$ spectroscopy.

In the standard AdS/CFT prescription, baryonic states, created by boundary operators like $\hat{\mathcal{O}}\sim \epsilon_{ijk}\, \hat{q}_i \,\hat{q}_j\,\hat{q}_k$ with $\text{dim}\,\bar{\mathcal{O}}=\Delta$, are dual to bulk normalizable fermionic modes that scale as $z^\Delta$. This \emph{matching} is a consequence of the so-called field/operator duality. These bulk fields obey the dynamics governed by action densities such as \eqref{Dirac-ac} and \eqref{rarita}. The information regarding the \emph{hadronic identity}, i.e., the dimension of $\bar{\mathcal{O}}$, is condensed in the fermionic bulk mass $M_5$. In general, for a given hadron, we can write the conformal dimension as a combination of the contribution from constituent quarks (twist) and the total orbital angular momentum $L$ as

\begin{equation}
 \Delta_{baryon}=\Delta_{q}+L.
\label{eq:o}
\end{equation}
In the case of baryons, the expression above becomes handy to deal with high fermionic spin. Higher spin contributions can be written with $L$. For $\Sigma$ with spin 1/2 and 3/2, their $L$  are 0 and 1, respectively.

Since a baryonic state has three quarks, each with a scaling dimension of $3/2$,  the constituent scaling dimension is $\Delta_{q}=9/2$. By substituting these data into Eq.\eqref{eq:o}, we can easily obtain the conformal dimensions $\Delta_{1/2}=9/2$ and $\Delta_{3/2}=11/2$ for the $\Sigma$ trajectories.

From the AdS/CFT dictionary,  we find the following relationship for the fermion bulk mass $M_{5}$ and its baryon conformal dimension, given by:

\begin{equation}
\left|M_5\right|=\Delta_{baryon}-2.
\label{eq:m5}
\end{equation}
Therefore, according to Eq. \eqref{eq:o}, for the $\Sigma$ baryons we obtain $M_{5}=5/2$ and $M_{5}=7/2$ for $\Sigma(1/2)$ and $\Sigma(3/2)$ respectively.

\subsection{Non-quadratic deformed background}
This model is a variation of the proposal presented in \cite{FolcoCapossoli:2019imm}, where quadratic deformations of the AdS warp factor, in Poincare coordinates, describe light hadrons. For strange baryons, we propose  $h(z)=\frac{1}{2}\,\left(k\,z\right)^{2-\alpha}$ and fix the dilaton to be zero. Thus, for the warp factor in the AdS metric, we can write the following expression.

\begin{equation}
    A\left(z\right)=\log \left(\frac{R}{z}\right)+\frac{1}{2}\,\left(k\,z\right)^{2-\alpha}.\label{eq:AZ1}
\end{equation}

The effect of the geometric deformation $h(z)$ is to place confinement. For fermions in the AdS space, bulk modes are unbounded. However, the deformation $h(z)$ causes the emergence of bounded states dual to baryons at the conformal boundary. Since we choose $\Phi(z)=0$, the bulk mass reduces to the standard value of $M_5$ given by Eq. \eqref{eq:m5}.

After these definitions, we obtain the  \emph{Schrödinger-like} equation for both the right and left bulk fermions as:

\begin{multline}
-\phi_{R/L}^{\prime \prime}(z)+\left[M_5^2 \,e^{2\,A(z)} \pm M_5 \,e^{A(z)} A^{\prime}(z)\right] \phi_{R/L}(z)\\
=m^2_n\,\phi_{R/L}(z), \label{eq:first}
\end{multline}
where $m_{n}$ is the four-dimensional fermion mass.

We fix the parameters in geometric deformation  $h(z)$, given in Eq. \eqref{eq:AZ1} as $k_{1/2}=239$ MeV, $k_{3/2}=219$ MeV and $\alpha_{1/2}=\alpha_{3/2}=0.16$ for each baryonic trajectory. By entering these parameters in Eq.\eqref{eq:first}, we calculated a baryonic mass spectrum that is consistent with $\Sigma(1/2)$ and $\Sigma(3/2)$ trajectories, as indicated in Tables \ref{table:table1} and \ref{table:table2}. The \emph{relative error} is presented in the last columns of Tables \ref{table:table1} and \ref{table:table2}. This error is calculated as follows

\begin{equation}
\% M=\sqrt{\left(\frac{\delta O_i}{O_i}\right)^2} \times 100,
\label{eq:M}
\end{equation}
where $\delta O_i$ depicts the deviations between the data $(M_\text{Exp})$ and the model prediction $(M_\text{Th})$.

\begin{table}[h!]
  \begin{center}
    \begin{tabular}{l c c c}
    \hline
    \hline
      \multicolumn{4}{c}{\textbf{$I(J^P)=1(1/2^+)$  states}}\\
      \hline
      \hline $n$&$M_\text{{th}}$(MeV)&$M_\text{Exp}$(MeV)&$\% M$\\
      \hline
      1&1135.22&1192.6$\pm$0.02&4.81\\
      2&1431.32&1585$\pm$20&9.70\\
      3&1717.22&1820 to 1940&8.66\\
      \hline
    \end{tabular}
    \caption{Masses of the $\Sigma(1/2)$ trajectory using the non-quadratic deformed background with $k_{1/2}=239$ MeV and $\alpha_{1/2}=0.16$. The ground state is represented by $n=1$. The last column is the relative error Eq. \eqref{eq:M} defines. For the mass intervals, we choose the average between the interval extremes. The experimental masses are read from the particle data group \cite{Workman:2022ynf}. }
    \label{table:table1}
  \end{center}
\end{table}

\begin{table}[h!]
  \begin{center}
    \begin{tabular}{l c c c}
      \hline
      \hline
      \multicolumn{4}{c}{\textbf{$I(J^P)=1(3/2^+)$  states}}\\
      \hline
      \hline $n$&$M_\text{th}$(MeV)&$M_\text{Exp}$(MeV)&$\% M$\\
      \hline
      1&1401.37&1382.83$\pm$0.34&1.28\\
      2&1675.77&1730 to 1830&5.86\\
      3&1942.61&1920 to 1960&0.13\\
      4&2203.36&2060 to 2120&5.42\\
      5&2459.06&2240$\pm$27&10.07\\
      \hline
    \end{tabular}
    \caption{Masses of $\Sigma(3/2)$ trajectory using the non-quadratic deformed background with $k_{3/2}=219$ MeV and $\alpha_{3/2}=0.16$. The ground state is represented by $n=1$. The last column is the relative error Eq. \eqref{eq:M} defines. For the mass intervals, we choose the average between the interval extremes. The experimental masses are read from the particle data group \cite{Workman:2022ynf}.}
    \label{table:table2}
  \end{center}
\end{table}

\begin{figure}
    \centering    \includegraphics[width=8.5cm]{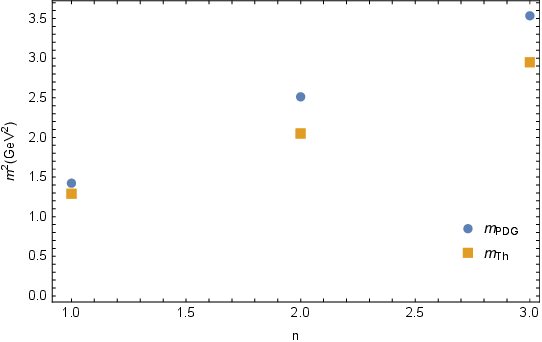}    \caption{\label{fig1} The $(n,m^2)$ Chew-Frautschi plot depicting the Regge trajectory for $\Sigma(1/2)$ baryon system computed from the  non-quadratic deformed background.}
\end{figure}

\begin{figure}
    \centering
    \includegraphics[width=8.5cm]{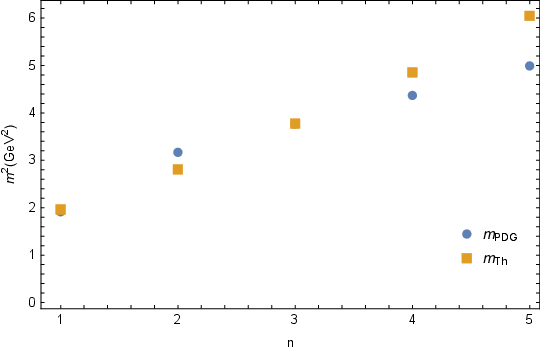}
    \caption{\label{fig2} The $(n,m^2)$ Chew-Frautschi plot depicting the Regge trajectory for $\Sigma(3/2)$ baryon system computed from non-quadratic deformed background.}
\end{figure}

We constructed $(n, m^2)$ \emph{Chew-Frautschi plots} using data from Tables \ref{table:table1} and \ref{table:table2}, which are presented in Figs.\ref{fig1} and \ref{fig2}. The figures show that the discrepancies between the calculated baryon masses and the experimentally measured mass data are below 10$\%$. Particularly noteworthy is the nearly identical masses for $\Sigma(1/2)$ with $n=1$, exhibiting an error of $4.8\,\%$, while for $\Sigma(3/2)$ with $n=3$, the error is a strikingly low 0.13$\%$.

 Additionally, we calculate the overall RMS error using the following definition:

\begin{equation}
\delta_{\text{RMS}}=\sqrt{\frac{1}{N-N_p} \sum_{i=1}^N\left(\frac{\delta O_i}{O_i}\right)^2} \times 100, \label{eq:totalrms}
\end{equation}

\noindent where $N$ and $N_p$ are the number of measurements and parameters, respectively. In this case, we obtain the values of the parameters $\alpha$ and $\kappa$ for the $\Sigma(1/2)$ and $\Sigma(3/2)$ trajectories separately by fitting and assigning a common value $\alpha$ for both trajectories. This choice comes from the Bethe-Salpeter analysis of Regge-trajectories \cite{Chen:2018bbr}, where the constituent quark mass influences the linearity of the trajectory. Since both $\Sigma$ trajectories have the same quark content, it is natural to consider the same $\alpha$ for them. From Eq.\eqref{eq:totalrms}, we find that $\delta_{\text{rms}}=8.5\,\%$ for $\Sigma$ trajectories using the deformed background.

\subsection{Non-quadratic dilaton}
This proposal is an extension of the non-quadratic dilaton idea developed for isovector mesons and non-$q\bar{q}$ states \cite{MartinContreras:2020cyg, MartinContreras:2023oqs}. In the case of fermionic bulk fields, the dilaton field factors out from the equations of motion. However, confinement is addressed by inserting the dilaton an \emph{anomalous dimension}, as we explain in section \ref{sec-2}.

In this prescription, we will set $h(z)=0$, and then the warp factor $A(z)$ and the dilaton field $\Phi(z)$ can be written as

\begin{eqnarray}
A(z)&=&\log \left(\frac{R}{z}\right),\\
\Phi(z)&=&\frac{1}{R}(\kappa\,z)^{2-\alpha}.
\end{eqnarray}

Then, using Eq.\eqref{eq:m5D}, we obtain a \emph{Schrödinger-like} equation:

\begin{multline}
   -\phi_{R / L}^{\prime \prime}(z)+\left\{\left[M_5+\Phi(z)\right]^2 \mathrm{e}^{2 A(z)} \pm \right.\\
   \left.\left[(M_5+\Phi) \mathrm{e}^{A(z)}\right]^{\prime}\right\} \phi_{R / L}(z)=m^2_n\, \phi_{R / L}(z). \label{eq:second}
\end{multline}

Notice that the bulk mass $M_{5}$ used here remains the same as in the deformed case. We set $\kappa_{1/2}=\kappa_{3/2}=0.423$ and $\alpha_{1/2}=\alpha_{3/2}=0.17$, and substitute them into Eq.\eqref{eq:second}. The resulting data is displayed in Tables \ref{table:table3} and \ref{table:table4}.

\begin{table}[h!]
  \begin{center}
    \begin{tabular}{l c c c}
      \hline
      \hline
      \multicolumn{4}{c}{\textbf{$I(J^P)=1(1/2^+)$  states}}\\
      \hline
      \hline
      $n$&$M_{th}$(MeV)&$M_{PDG}$(MeV)&$\% M$\\
      \hline
      1&1377.51&1192.6$\pm$0.02&15.50\\
      2&1555.77&1585$\pm$20&1.84\\
      3&1712.28&1820 to 1940&8.92\\
      \hline
    \end{tabular}
    \caption{Mass spectrum for the $\Sigma(1/2)$ trajectory within the non-quadratic dilaton model. We use $\kappa=0.423$ MeV and $\alpha=0.17$. The ground state is represented by $n=1$. As customary, $\%M$ is the relative error defined by Eq. \eqref{eq:M}. For the mass intervals, we choose the average between the interval extremes. The experimental masses are read from PDG \cite{Workman:2022ynf}.}
    \label{table:table3}
  \end{center}
\end{table}

\begin{table}[h!]
  \begin{center}
    \begin{tabular}{l  c c c}
        \hline
        \hline
      \multicolumn{4}{c}{\textbf{$I(J^P)=1(3/2^+)$  states}}\\
      \hline
      \hline
      $n$&$M_{th}$(MeV)&$M_{PDG}$(MeV)&$\% M$\\
      \hline
     1&1571.94&1382.83$\pm$0.34&13.60\\
      2&1726.75&1730 to 1830&3.00\\
      3&1866.39&1920 to 1960&3.79\\
    4&1994.45&2060 to 2120&4.57\\
      5&2113.28&2240$\pm$27&5.40\\
      \hline
    \end{tabular}
    \caption{Mass spectrum of $\Sigma(3/2)$ trajectory  within the non-quadratic dilaton model. We set $\kappa=0.423$ MeV and $\alpha=0.17$. The ground state is represented by $n=1$. As customary, $\%M$ is the relative error defined by Eq.\eqref{eq:M}. For the mass intervals, we choose the average between the interval extremes. The experimental masses are read from PDG \cite{Workman:2022ynf}.}
    \label{table:table4}
  \end{center}
\end{table}

 We have generated the $(n, m^2)$ Chew-Frautschi plot based on the data presented in Tables \ref{table:table3} and \ref{table:table4}. The non-quadratic dilaton accurately matches the high excited states: the relative errors are below $6\,\%$. However, the ground state is not well-fitted. It is unsurprising since the ground state strongly depends on the dilaton slope $\kappa$.

\begin{figure}
    \centering
    \includegraphics[width=8.5cm]{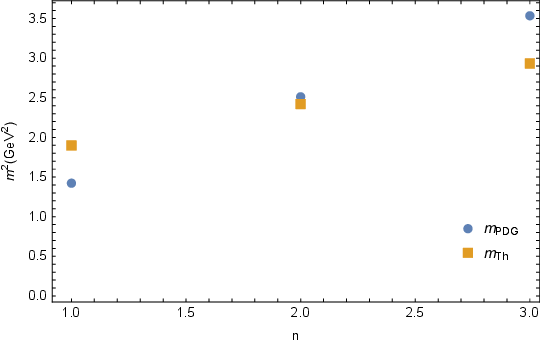}
    \caption{\label{fig3} The $(n,m^2)$ Chew-Frautschi plot for the $\Sigma(1/2)$ trajectory using the non-quadratic dilaton.}
\end{figure}

\begin{figure}
    \centering
    \includegraphics[width=8.5cm]{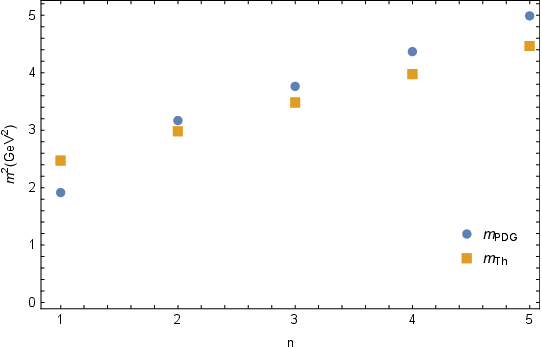}
    \caption{\label{fig4} The $(n,m^2)$ Chew-Frautschi plot for $\Sigma(3/2)$ trajectory using the non-quadratic dilaton.}
\end{figure}

Lastly, we computed the total RMS error, and it is important to mention that we employed identical parameters for $\Sigma$ with different spins. Therefore, $N_{p}=2$ is involved in this calculation. From Eq.\eqref{eq:totalrms}, we find $\delta_{\mathrm{rms}}=9.84\,\%$ for the $\Sigma$ family.

\section{Configurational Entropy}\label{sec-4}

\subsection{Configurational entropy and hadronic stability}

The inspiration behind Configuration Entropy \cite{Gleiser:2011di, Gleiser:2012tu, Gleiser:2013mga} comes from Shannon's information entropy \cite{6773024}, a measure that quantifies the amount of information in a message \cite{Braga:2020opg}. For a variable that can take $N_{d}$ discrete possible values, with probabilities given by $p_{i}$, it is defined by

\begin{equation}
    S_{\text {info }}=-\sum_{i=1}^{N_d} p_i \ln p_i.
\end{equation}


In the original formulation within information theory, configurational entropy (CE) can be understood as a measure of the information required to describe localized functions or sets of solution parameters. Generally, dynamical solutions emerge from optimizing an action, and configurational entropy measures the information available in these solutions. The relationship between configurational entropy and complexity is associated with stability. As configurational entropy characterizes the complexity of a given physical system, states with higher configurational entropy require more energy for their occurrence in nature compared to states with lower configurational entropy. Higher energy levels also imply more modes that conform to such physical states, indicating that configurational entropy increases with higher coarseness. Hence, configurational entropy can also be perceived as a measure of stability. Remember that configurational entropy reflects the relative ordering in field configuration space, illustrating the relationship between energy and coarseness. With an increase in constituents, there is a corresponding increase in energy and relative configurational entropy. Additionally, investigations on active matter systems demonstrate that the final state of a many-particle system minimizes configurational entropy when it reaches equilibrium.

Configuration Entropy is commonly used to describe the variety of particle spatial arrangements within a system at the microscopic scale. Its calculation involves determining the quantity and distribution of distinct microscopic states throughout the system. Specifically, CE measures the complexity and diversity within a system, reflecting the uncertainty associated with particle arrangements at the microscopic level. When computing the configurational entropy for the $\Sigma$ family, from a bottom-up perspective, we connect bulk localization of the dual fermionic modes with the stability at the boundary. This fact is connected with the observation that heavier states should decay faster than light ones. Thus, we expect that the CE decreases with the excitation level $n$.

Thus, we can say that holographic CE indirectly measures the diversity and complexity of constituent spatial arrangements in the $\Sigma$ baryon family. A high configurational entropy in the $\Sigma$ system implies that the microscopic particle arrangement exhibits a high degree of randomness and disorder, potentially making the system more susceptible to decay or fragmentation, thus reducing its stability. In contrast, a low configurational entropy suggests a more compact and orderly microscopic particle arrangement, often indicative of a more stable system.

In recent years, numerous studies have focused on configurational entropy, including investigations into compact objects \cite{Gleiser:2015rwa} and holographic AdS/QCD models \cite{Bernardini:2016hvx, Bernardini:2016qit, Braga:2016wzx, Braga:2017fsb, Braga:2018fyc, Bernardini:2018uuy, Ferreira:2019inu, MartinContreras:2023eft}, among other diverse systems. These studies, approached from various angles, consistently demonstrate a parallel relationship between changes in configurational entropy and stability. In this paper, we perform holographic calculations of the configurational entropy for baryons using different holographic models discussed in the previous section.

To do so, we compute the differential configurational entropy (DCE) for a given physical system by performing the following calculations for each model: Firstly, we obtain the localized solutions to the equations of motion. Then, we evaluate the on-shell energy density. Next, we transform the on-shell energy density into momentum space. Finally, we calculate the modal fraction and evaluate the DCE integral based on the results.

\subsection{Configurational entropy for baryons}

The key ingredient for DCE comes from the on-shell energy-momentum tensor for the bulk fields. Since the AdS frame is static to these fields, it is possible to use the \emph{dynamical}
On-shell energy tensor is defined as

\begin{equation}
T_{mn}=-\frac{2}{\sqrt{-g}}\frac{\delta\,I}{\delta\,g^{mn}}.
\end{equation}

For the DCE, it is enough to consider this definition since gravity has not been affected by the fermionic fields. However, the variations must be done using the frame fields $e^a_m$ instead of the metric tensor. Thus, we will follow the prescription in Ref. \cite{Shapiro:2016pfm}. Therefore, from the action \eqref{Dirac-ac} we obtain for spin $1/2$

\begin{equation}
    T_{mn}=\frac{1}{2\,\mathcal{K}}\bar{\psi}\left(\Gamma_m\,\overset{\leftrightarrow}{D}_n+\Gamma_n\,\overset{\leftrightarrow}{D}_m\right)\psi,
\end{equation}

\noindent and for spin $3/2$, from action \eqref{rarita}, we have:

\begin{multline}
    T_{mn}=-\frac{1}{2\,\mathcal{K}}\,\bar{\psi}^{r p}\left(\Gamma_m\,\overset{\leftrightarrow}{\nabla}_n+\Gamma_n\,\overset{\leftrightarrow}{\nabla}_m\right)\psi_{rp}\\
    +\frac{1}{\mathcal{K}}\left[\bar{\psi}_m^{p}\,\Gamma^r\,\overset{\leftrightarrow}{\nabla}_r\,\psi_{n\, p}\right.\\
    \left.-M_5(z)\,\bar{\psi}_m^{l}\psi_{n l}\right].
\end{multline}

The energy density in both cases is extracted from the $T_{00}$ component. For the spin $1/2$ field, after Fourier transform, we find

\begin{equation}
    \rho_{1/2}(z)=\frac{1}{\mathcal{K}}e^{A(z)}\,m_n\,\left(\phi_{L,n}^2+\phi_{R,n}^{2}\right)\,\mathcal{A}_0,
\end{equation}

\noindent where $\mathcal{A}_0$ is a polarization factor. The energy density for the spin $3/2$ field is

\begin{multline}
\rho_{3/2}(z)= -\frac{1}{\mathcal{K}}\,e^{-A(z)}\left[m_n\left(2\,\mathcal{A}_1-\mathcal{A}_2\,e^{2\,A(z)}\right)\right.\\\
      \left.+\,M_5\,\mathcal{A}_2\,e^{2\,A(z)}\right]\left(\phi_{L,n}^2+\phi_{R,n}^{2}\right),
\end{multline}

\noindent where $\mathcal{A}_1$ and $\mathcal{A}_2$ are polarization factors that appear from the contraction of the indices. For simplicity, we will choice $\mathcal{A}_1=\mathcal{A}_2=\mathcal{A}_0$. Then, we perform a Fourier transform on the energy density and express it as

\begin{equation}
    \bar{\rho}(k)=\int_0^{\infty} d \zeta e^{- i k \zeta} \rho(\zeta).
\end{equation}

The modal fraction, which describes how localized the information is in a given mode,  is defined as

\begin{equation}
    f(k)=\frac{|\bar{\rho}(k)|^2}{\int d k|\bar{\rho}(k)|^2}.
\end{equation}

For the continuous variables case, we use the differential configurational entropy (DCE) defined as

\begin{equation}
    S_{D C E}=-\int d k \tilde{f}(k) \log \tilde{f}(k),
\end{equation}

\noindent where $\tilde{f}(k)=f(k) / f(k)_{\operatorname{Max}}$, and $f(k)_{\operatorname{Max}}$ is the maximum value assumed by $f(k)$.

Recall that $\rho(z)\in L^2(\mathbb{R})$ has information on how energy is localized in the bulk. Thus, it indirectly measures how normalizable modes are well localized in the AdS space. In hadronic terms, this localization measure is also a signal of confinement since a bounded state should be localized. Therefore, DCE becomes a clear test for holographic models mimicking hadrons.

\begin{center}
\begin{figure}
\includegraphics[width=8.5cm]{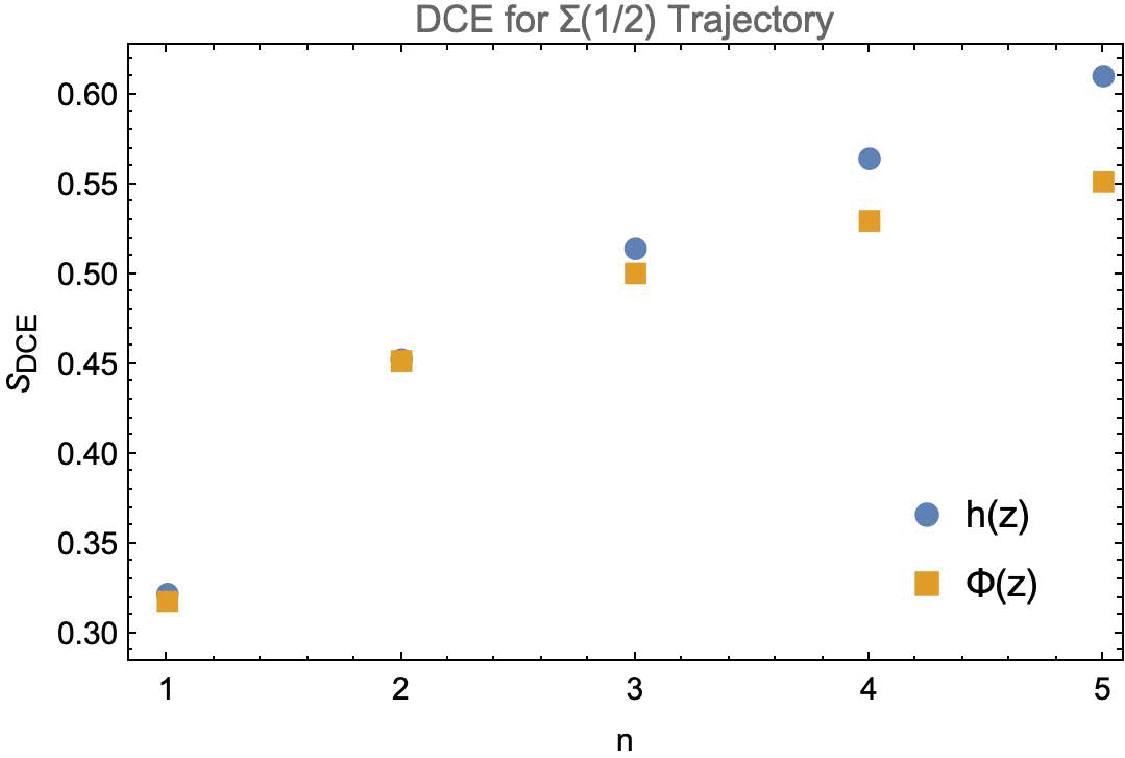}
\caption{\label{fig5} The differential configurational entropy (DCE) for the two holographic models describing the $\Sigma(1/2)$ trajectory as a function of excitation number $n$. Blue circles represent non-quadratic geometric deformation, and orange squares represent non-quadratic dilaton.}
    \label{fig:one}
\end{figure}
\end{center}

\begin{center}
    \begin{figure}
    \includegraphics[width=8.5cm]{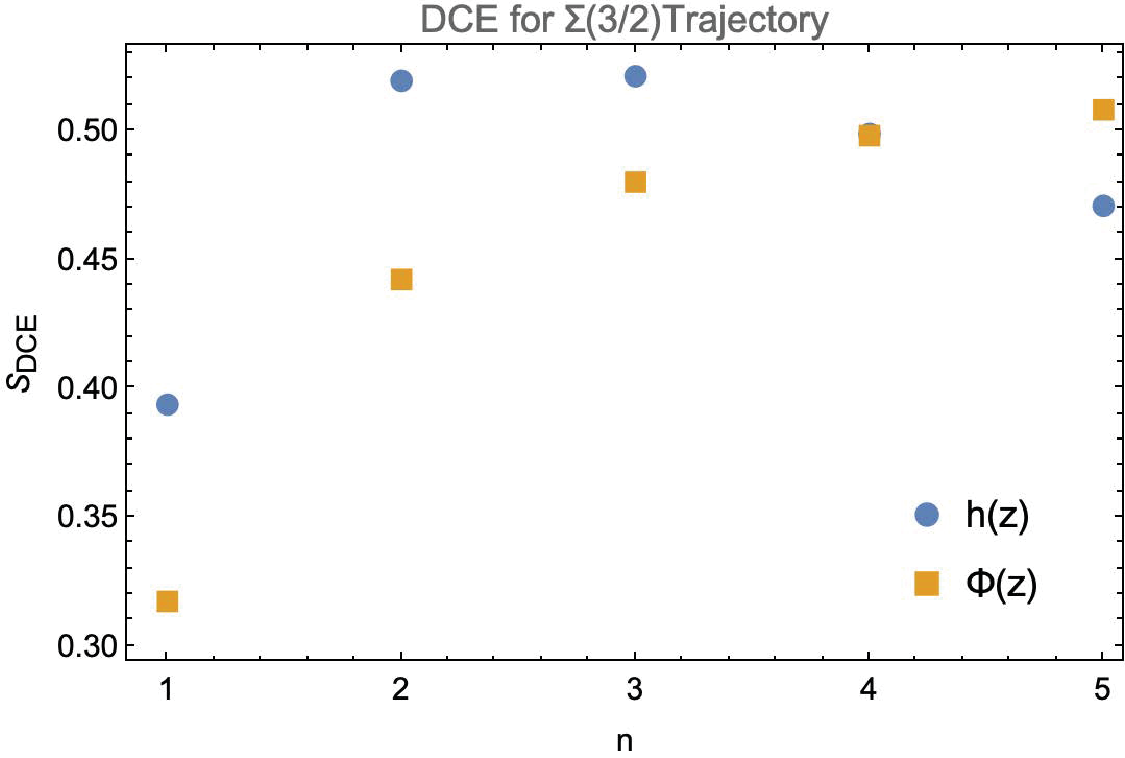}
    \caption{\label{fig6} The differential configurational entropy (DCE) for the two holographic models describing the $\Sigma(3/2)$ trajectory as a function of the excitation number $n$. Blue circles represent non-quadratic geometric deformation, and orange squares represent non-quadratic dilaton.}
        \label{fig:enter-label}
    \end{figure}
\end{center}

We calculated the differential configurational entropy (DCE) for the $\Sigma(1/2)$ and $\Sigma(3/2)$ trajectories using both bottom/up approaches. We summarize our findings in Figs.\ref{fig5} and \ref{fig6}.

As we expected from the localization and stability hypothesis, Fig. \ref{fig5} shows that as $n$ increases, $S_\text{DCE}$ increases for the $\sigma(1/2)$ trajectory in both models. However, for the $\Sigma(3/2)$ trajectory, the DCE only grows with $n$ for the non-quadratic dilaton approach. The deformed background approach increases only for the first excited state. Higher excitations have decreased the DCE. Assuming the DCE-stability hypothesis, these higher states \emph{become more stable}. This conclusion is not possible from the hadronic phenomenology. Thus, \emph{deformed background approach} is not good for describing these $\Sigma(3/2)$ states.

\section{Conclusions}\label{sec-5}
In this study, we developed two models: the non-quadratic deformed background with the warp factor set as $A\left(z\right)=\log \left(\frac{R}{z}\right)+\frac{1}{2}\,\left(k\,z\right)^{2-\alpha}$. For the second model, we assume a non-quadratic dilaton $\Phi(z)=\frac{1}{R}(\kappa\,z)^{2-\alpha}$. Using standard bottom-up techniques, we tested these models as alternatives to describe $\Sigma$ baryon spectroscopy. Each model has two parameters associated with the Regge slope and the linearity deviation. In both cases, experimental data was properly fitted, having RMS errors smaller than 10 $\%$. The next stage in choosing a good model is the DCE analysis.

Regarding the mass spectrum, we observed that ground states are not well-fitted in both models. In other bottom-up models, ground state mass sets the Regge slope. However, in this non-quadratic scenario, the slope and the deformation are fitted by regression over the entire trajectory. Ground states are strongly attached to the behavior of the one-gluon exchange term in hadronic potentials. These terms come from the perturbative analysis, which is not captured in the dilaton or deformations. Recall that this bottom-up confinement tries to mimic the confinement part in the Cornell-like potentials that control the higher excitations. Further improvements in the determination of the intercept (leading to a better ground state mass) are required. For further details, see  \cite{Afonin:2021cwo} for an interesting discussion and review of how hadronic spectroscopy can be captured in bottom-up models.

Configurational entropy measures how well localized a mode is in the solution space. Thus, it could have information about confinement by considering that the emergence of bounded colorless states is a consequence of color confinement. Thus, DCE can be used to address whether or not a given holographic approach is suitable to describe hadrons from the stability point of view. However, DCE is not the only test we have. Thermal analysis \cite{MartinContreras:2021yfz} of the two-point spectral function also discusses stability.

For the models discussed here, we observed that the deformed background seems inconsistent with the hypothesis of stability/DCE, at least for the $\Sigma(3/2)$ trajectory. As a hypothesis, if we analyze the behavior of the Schrödinger modes in the deformed background, we see that they are highly suppressed in the bulk due to the structure of the holographic potential, i.e., the exponential factor in Eqn. \eqref{fermion-pot-sch}, the modes are spatially confined. For excited modes, they oscillate in small bulk regions. Since these solutions do not tend to smear out into the bulk, the DCE does not decrease. This hypothesis has to the tested with other similar geometric approaches.

From its original motivation, differential configurational entropy describes how constituents are distributed among different states or configurations in a given system. It allows us to quantify the degree of disorder or randomness of those constituent arrangements, providing insights into the probabilities of various particle configurations. Thus, DCE is related to how constituents interact with each other. This fact allows DCE to be considered a probe to test how confinement is realized in holographic models. However, there is still plenty of room to discuss the nature of the arrangements of constituents inside hadrons since how we model these systems in such holographic models has to be improved.

A higher differential configurational entropy indicates more possible particle arrangements, implying greater disorder or freedom of movement. On the other hand, a lower differential configurational entropy suggests a more ordered particle arrangement or more significant constraints on the configuration. The calculation results for the two models are shown in Figs. \ref{fig5} and \ref{fig6}. From the figures, we can observe that the data overall increases with the increase in mass, and the values of configurational entropy also increase. By comparing the results of the two models, it is found that the model incorporating the dilaton field obtains lower configurational entropy in the calculation of $\Sigma (1/2)$ and $\Sigma (3/2)$.

The relationship between configurational entropy and the mass spectrum of $\Sigma$ baryons is highly significant. Configurational entropy is an indirect measure of the intricacy of the structure in $\Sigma$ baryons. Meanwhile, the mass spectrum characterizes the distribution of masses corresponding to different energy levels and combinations of constituent particles within $\Sigma$ baryons. Higher configurational entropy in $\Sigma$ baryons signifies a broader range of arrangements and degrees of freedom. This observation can also be interpreted in terms of transition probabilities. Higher DCE is connected with smaller decay widths. Thus, the smallest DCE is expected to belong to the ground state with the highest decay width. We will explore this idea in further investigations.

\begin{acknowledgments}
M. A. Martin Contreras wants to acknowledge the financial support provided by the National Natural Science Foundation of China (NSFC) under grant No 12350410371. X. Chen wants to acknowledge the financial support provided by Research Foundation of Education Bureau of Hunan Province, China (Grant No. 21B0402) and the Natural Science Foundation of Hunan Province of China under Grant No.2022JJ40344.
\end{acknowledgments}
\bibliography{apssamp}
\end{document}